\documentclass[a4paper,11pt]{article}
\usepackage{jcappub} 
\usepackage{lineno}
\usepackage{mathtools}

\title{\boldmath New physics as a possible explanation for the Amaterasu particle}

\author{Rodrigo Guedes Lang}
\affiliation{Erlangen Centre for Astroparticle Physics, Friedrich-Alexander-Universität Erlangen-Nürnberg, \\ Nikolaus-Fiebiger-Str. 2, Erlangen, D-91058, Germany}

\emailAdd{rodrigo.lang@fau.de}

\abstract{The Telescope Array experiment has recently reported the most energetic event detected in the hybrid technique era, with a reconstructed energy of 240 EeV, which has been named ``Amaterasu'' after the Shinto deity. Its origin is intriguing since no powerful enough candidate sources are located within the region consistent with its propagation horizon and arrival direction. In this work, we investigate the possibility of describing its origin in a scenario of new physics, specifically under a Lorentz Invariance Violation (LIV) assumption. The kinematics of UHECR propagation under a phenomenological LIV approach is investigated. The total mean free path for a particle with Amaterasu's energy increases from a few Mpc to hundreds of Mpc for $-\delta_{\rm{had},0} > 10^{-22}$, expanding significantly the region from which it could have originated. A combined fit of the spectrum and composition data of Telescope Array under different LIV assumptions was also performed. The data is best fitted with some level of LIV both with and without Amaterasu. Robustness with data from the Pierre Auger Observatory is investigated by exploring an intermediate composition scenario. Similar improvements in the description of the data with LIV are found for that. New physics in the form of LIV could, thus, provide a plausible and robust explanation for the Amaterasu particle.}

\begin{document}
\maketitle
\flushbottom

\section{\label{sec:intro}Introduction}

Ultra-high-energy cosmic rays (UHECR, $E > 10^{18}$~eV) are the most energetic particles known in the Universe, forming the high-energy end of the energy spectrum of detected astroparticles. Due to their charged nature, which leads to deviations during the propagation in the presence of magnetic fields, their origins remain an open puzzle. Recently, an extra piece was added to this puzzle with the report of Telescope Array's detection of an UHECR air shower with a reconstructed energy of $(244 \pm 29 , \rm{(stat.)} ^{+51}_{-76} \, \rm{(syst.)})$~EeV (about 40 Joules) and reconstructed direction pointing to $(255.9 \pm 0.6, \, 16.1 \pm 0.5)^{\circ}$ in equatorial coordinates~\citep{Amaterasu}.

This unique event was named Amaterasu after the Shinto goddess of the sun and represents the second-highest nominal energy ever detected (and highest in the hybrid detection technique era), only after the famous ``Oh-My-God'' particle detected with $(320 \pm 0.9)$~EeV by the Fly Eye's experiment in 1991~\citep{OMG}. Moreover, its arrival direction is intriguing. UHECR with such energy are expected to quickly lose energy due to interactions with background photons, leading to a propagation horizon of a few tens of Mpc~\citep{LangPRD1}. At the same time, at these energies, the magnetic field deviations are expected to be of the order of $20^{\circ}$ in the worst-case scenario~\citep{UngerFarrar}. With that, the origin point of Amaterasu could be restricted to a relatively small portion of the sky. Nevertheless, Amatarasu's direction points to the Local Void, a region of the sky with a relatively small number of galaxies ~\citep{Tully:2007ue}. Ref.~\citep{UngerFarrar} have further investigated it with the most up-to-date models for Galactic magnetic fields~\citep{PlanckGMF,Unger:2023lob}, finding no powerful enough candidate sources.

Possible astrophysical explanations have been proposed and include, e.g., transient events~\cite{Farrar:2024zsm}, UHECR clustering~\cite{universe10080323} and ultra-heavy UHECR~\cite{Zhang:2024sjp}. In another category of possible explanations, new physics could lead to scenarios in which Amaterasu can be described, in particular this is the case for Lorentz Invariance Violation (LIV). The breaking of Lorentz symmetry has either been proposed or accommodated by several beyond the Standard Model theories~\citep{Mattingly}. The effects of LIV are expected to be suppressed up to the highest energies and, thus, astroparticles, in particular UHECR, have been extensively explored as important tools for testing such theories in a phenomenological approach~\citep{LangUniverse,Scully:2008jp,Bi:2008yx,Maccione:2009ju,Lang:2019kA,AugerLIV}. 

In this work, we investigate the possibility of describing Amaterasu under a scenario with Lorentz Invariance Violation. In section~\ref{sec:LIV}, we describe the considered LIV framework as well as the corresponding effects on the propagation of UHECR. In section~\ref{sec:fit}, we present a combined fit for the data of Telescope Array under different LIV scenarios and how this is affected by Amaterasu. In section~\ref{sec:auger}, we explore the compatibility of the scenarios with results from the Pierre Auger Observatory. Finally, section~\ref{sec:conclusions} discusses this work's conclusions and final remarks.

\section{\label{sec:LIV}Lorentz Invariance Violation}

\subsection{Framework}

The most commonly used LIV framework in phenomenological studies reduces the main effects to a shift in the energy dispersion relation~\citep{ColemanGlashow1,ColemanGlashow2} (an extensive review on works using this and other formalisms can be found in Ref.~\cite{Addazi:2021xuf}),

\begin{equation}
\label{eq:LIV}
E^{2}_{a} = p^{2}_a + m^{2}_a + \sum_n \delta_{a,n} E^{n+2},
\end{equation}

where $E$, $p$ and $m$ denote the particle's energy, momentum, and mass, respectively. The particle species is denoted by $a$, given that LIV effects could in principle be independent for each particle type. The LIV effects of each order, $n$, are modulated by the LIV coefficient $\delta_{a,n}$, which can be either positive or negative. The most common approach is to investigate each LIV sector and order independently, i.e., to consider $\delta_{a,n} \neq 0$ for only a given $(a,n)$.

In this work, we use a similar approach to that used by the Pierre Auger Collaboration in their search for LIV imprints on their data~\citep{AugerLIV}. We consider LIV only in the hadronic sector and investigate only the leading order ($\delta_{\rm{had},0} \neq 0$). As in Refs.~\cite{Scully:2008jp} and \cite{AugerLIV}, the LIV coefficients of protons, nuclei and pions relate via

\begin{equation}
    \delta_{\rm{had},0} \coloneqq \delta_{p,0} = \frac{\delta_{\pi,0}}{2} = A^{n} \delta_{N,0}.
\end{equation}

While the leading order, $n=0$, is shown as the representative case, similar effects are expected for the higher orders. For the case of UHECR, the LIV coefficients from different orders can easily be obtained (eq. 6.4 of Ref. \cite{AugerLIV}).

\subsection{Propagation effects}

UHECR propagation is limited by interactions with background photons. At the highest energies, two main interactions are dominant. Protons and heavier nuclei lose energy due to photopion production ($p + \gamma \rightarrow p + \pi$) and nuclei undergo photodisintegration, emitting one or two nucleons and becoming a lighter nucleus, ($^{A}N + \gamma \rightarrow ^{A-1}N + p/n$). These interactions govern the observables on Earth, strongly changing what was accelerated at the sources. Both interactions are strongly energy dependent, becoming effective above a few tens of EeV. With that, an energy-dependent propagation horizon is created, leading to an effective spectral index on Earth which differs from that accelerated at the sources. The photodisintegration acts on changing the particle species, creating a larger number of lighter and lower energy UHECR from a single higher energy heavy UHECR. This strongly influences both the spectrum and composition measured on Earth~\citep{LangPRD1}.

\begin{figure}[ht]
\centering
\includegraphics[width=0.49\textwidth]{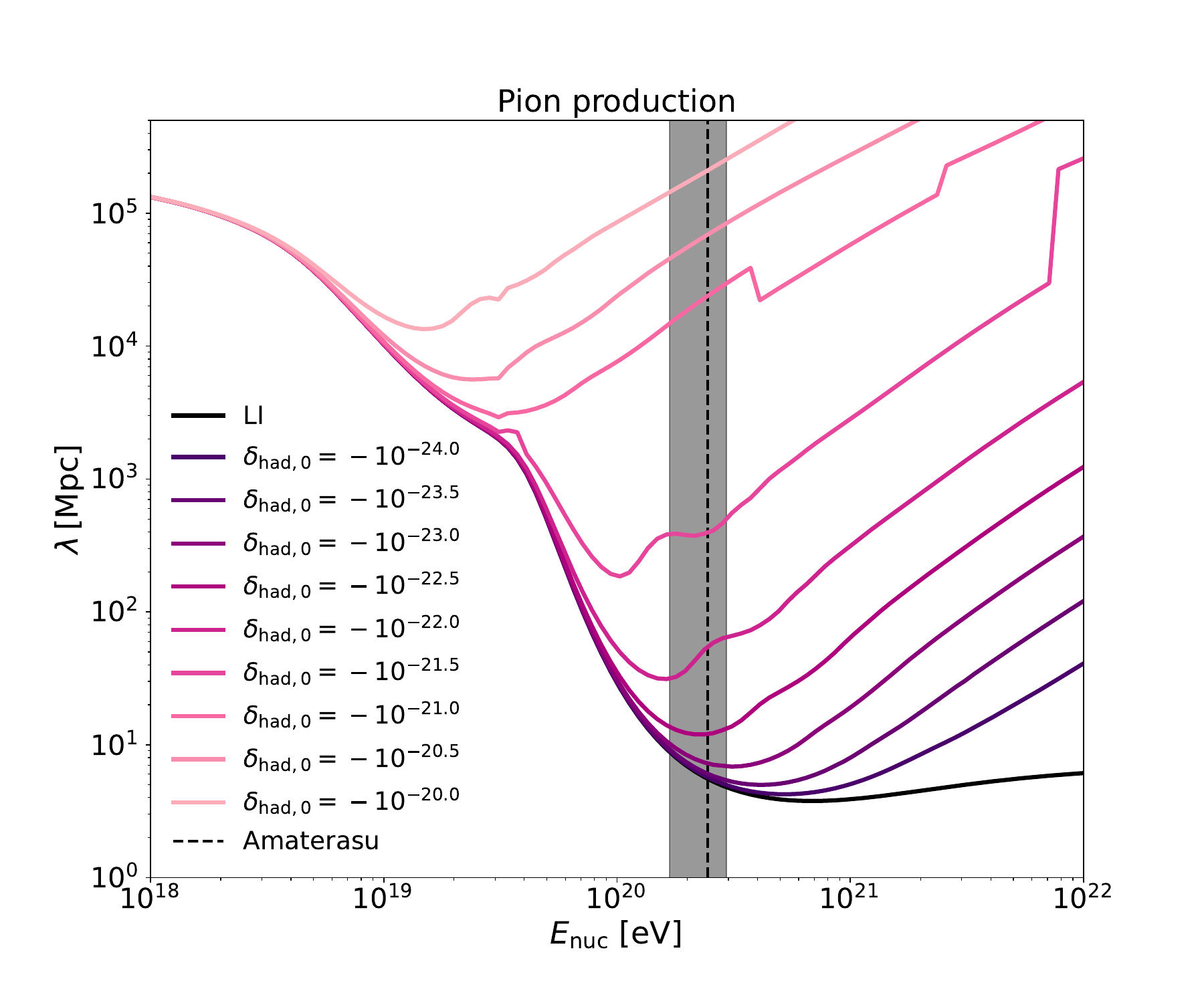}
\includegraphics[width=0.49\textwidth]{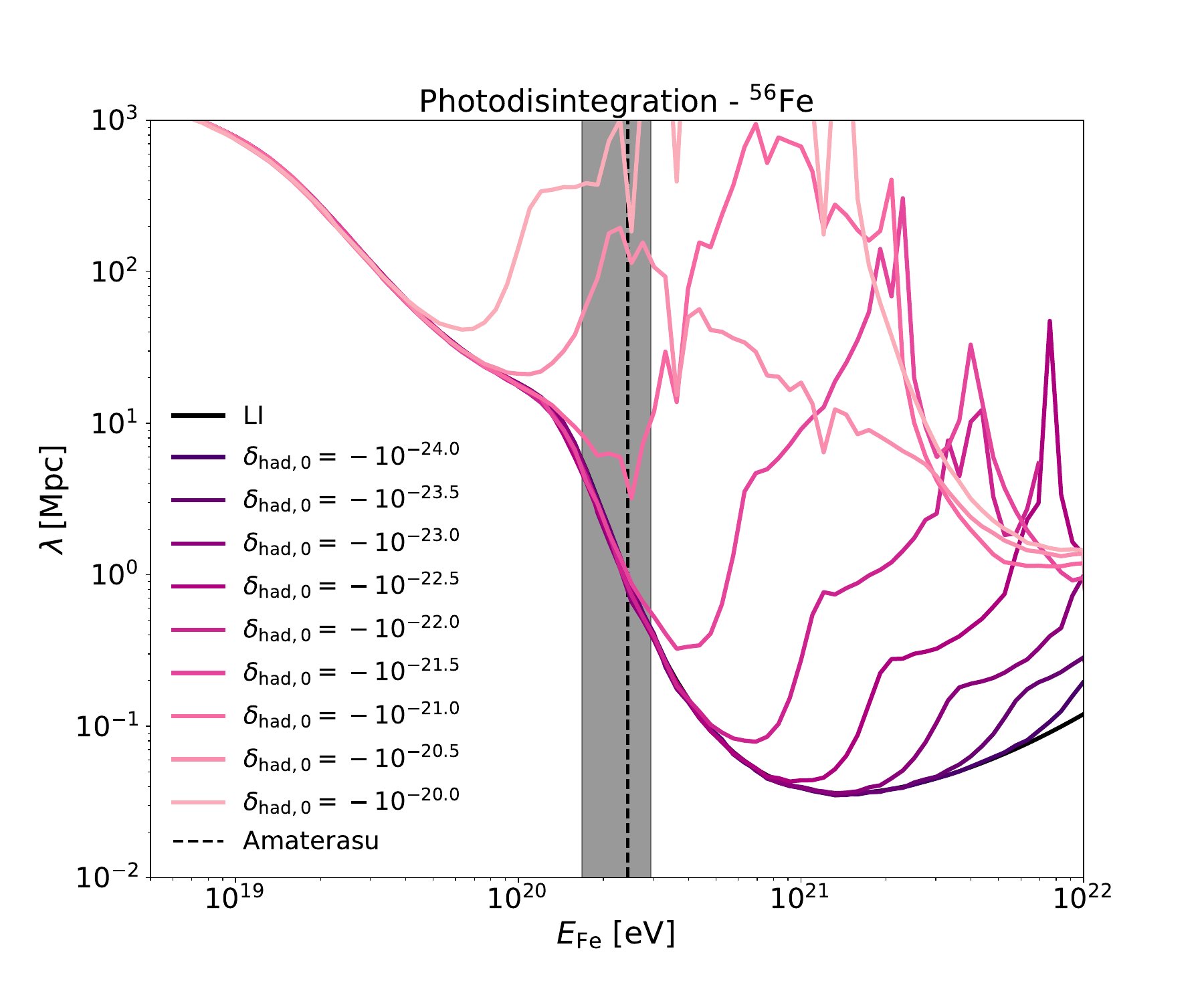}
\caption{\label{fig:LIVlambda}Mean free path as a function of the energy for different LIV scenarios. The different colors represent scenarios with increasing LIV coefficients. The vertical dashed line and shaded area show the detected energy for Amaterasu and the range of systematic uncertainties. The left and right panels are for photopion production and photodisintegration for an iron nucleus, respectively. The calculations consider the Gilmore model for the EBL background~\citep{Gilmore12}.}
\end{figure}

A change in the energy dispersion such as the one proposed in eq.~\ref{eq:LIV} would impact the kinematics of these interactions~\citep{Scully:2008jp,LangThesis,LangUniverse}. In this work, we use the calculations for the LIV kinematics proposed by~\cite{LangThesis} and \cite{LangUniverse}. The main effect for $\delta_{\rm{had},0} < 0$ is a reduction of the allowed four-momentum phase space of the particles, leading to a reduction in the number of interactions and, thus, an increase in the mean free path of propagating UHECR. This is illustrated in Fig.~\ref{fig:LIVlambda}, which shows the mean free path for both interactions under different LIV scenarios. The mean free path diverges from the LI case above a given energy, which is modulated by the absolute value of $\delta_{\rm{had},0}$.

\begin{figure}[t]
\centering
\includegraphics[width=0.8\textwidth]{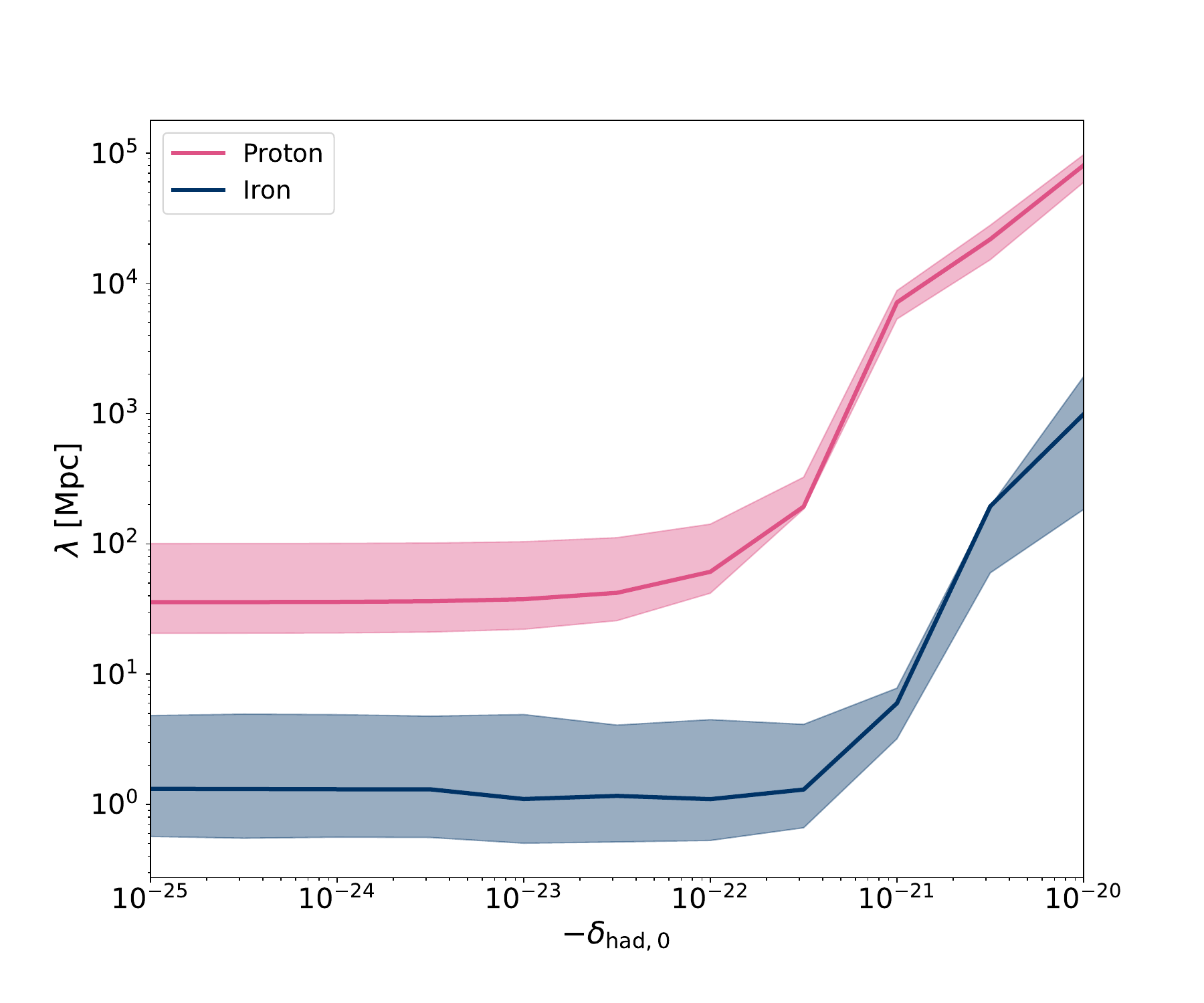}
\caption{\label{fig:LambdaVersusLIV}Mean free path as a function of the LIV coefficient for the detected energy of Amaterasu. The bracketing hypotheses of a primary proton and primary iron are considered. The shaded areas represent the results for different energies within the systematic uncertainties. The calculations consider the Gilmore model for the EBL background~\citep{Gilmore12}.}
\end{figure}

The total mean free path is particle dependent and, for nuclei, a sum of both effects. Fig.~\ref{fig:LambdaVersusLIV} shows how the total mean free path for a particle with Amaterasu's energy changes as a function of the LIV coefficient for the two bracketing hypotheses, i.e., a proton or an iron primary. A LIV coefficient of $-\delta_{\rm{had},0} > 10^{-22}$ is large enough to introduce a significant effect, increasing the mean free path up to three orders of magnitude for $\delta_{\rm{had},0} = -10^{-20}$. In those LIV scenarios, Amaterasu would be able to travel much farther without interacting and, thus, its acceleration site would not be required to be relatively close, solving, thus, the inconsistency with Amaterasu's large detected energy and reconstructed direction pointing to the Local Void. Similar effects on the mean free paths are found for higher LIV orders, $n>0$.

\subsection{Extensive air shower development effects}

A LIV scenario would affect not only interactions during UHECR propagation but also, e.g., interactions during the development of the extensive air shower (EAS) in the atmosphere. Investigations of LIV effects on EAS include an increase of the depth of the first interaction leading to a suppression of EAS formation~\cite{Satunin:2021vfx}, a shift in the $X_{\rm{max}}$ distribution due to photon and pion decay~\cite{Duenkel:2021gkq}, and changes on the muon content of EAS~\cite{PierreAuger:2021mve}. 

Nevertheless, a recent work by the Pierre Auger Observatory has constrained the LIV phase space in which effects are significant in the range currently explored by UHECR experiment to $\eta^{(1)} < -5.95 \times 10^{-6}$ at 90.5\% confidence level~\cite{PierreAuger:2021mve}. Since the same LIV formalism is used, the equivalent range for $\delta_{\rm{had},0}$ can be calculated:

\begin{equation}
    \delta_{\rm{had},1} = \delta_{\pi,1}/2 = \frac{\eta^{(1)}}{2 M_{\rm{Pl}}},
\end{equation}

and with eq. 6.4 of Ref.~\cite{AugerLIV},

\begin{equation}
   \delta_{\rm{had,0}} \sim \delta_{\rm{had},1} R_{\rm{max}},
\end{equation}

where $R_{\rm{max}}$ is the maximum rigidity accelerated at the sources. Even for the most conservative scenario considered in this work, i.e., $R_{\rm{max}} = 1$ EV, the resulting LIV coefficient is $\delta_{\rm{had,0}} \gtrsim - 2 \times 10^{-17}$. This is three orders of magnitude stronger than the most extreme LIV scenario considered in this work ($\delta_{\rm{had},0} = - 10^{-20}$). Therefore, the effects on EAS can safely be neglected in this analysis.

\section{\label{sec:fit}Combined fit under LIV assumptions}

If LIV is present, it should not be a unique characteristic of a single event but an effect perceived by every particle. For that reason, a LIV scenario that softens the tension about the origin of Amaterasu should still be consistent with the remaining of experimental data, i.e., it still requires an astrophysical model that describes under the same LIV assumption the other events measured by the Telescope Array experiment. Hence, we present a combined fit of the spectrum and composition data measured by the Telescope Array under different LIV scenarios.

\subsection{Dataset and fit method}

Following Ref.~\citep{TACombinedFit}, we use the 9-year Telescope Array energy spectrum~\citep{TASpectrum} and depth of shower maximum, $X_{\rm{max}}$, distributions~\citep{TAXmax}. The spectral counts are divided into 10 bins per decade of energy, ranging from $10^{18.8}$~eV to $10^{20.4}$~eV. The $X_{\rm{max}}$ counts are divided into 4 main energy bins ($18.8 < \log_{10} (E/\rm{eV}) < 19.0$, $19.0 < \log_{10} (E/\rm{eV}) < 19.2$, $19.2 < \log_{10} (E/\rm{eV}) < 19.6$ and $19.6 < \log_{10} (E/\rm{eV}) < 20.0$) and then into 21 linear bins of $X_{\rm{max}}$, ranging from 590 to 1010 $\rm{g/cm^{2}}$. Amaterasu does not influence the $X_{\rm{max}}$ counts as it was only detected with the surface array (SD).

The combined fit was based on a widely used procedure~\citep{TACombinedFit,AugerCombinedFit,AugerCombinedFit2,AugerLIV}. We simulate UHECR propagation using the state-of-the-art Monte Carlo package \texttt{CRPropa3}~\citep{CRPropa3}. The LIV effects were taken into account by implementing the modified mean free paths for each considered value of $\delta_{\rm{had},0}$ in the code. Sources were initially considered in a 3-D grid of energy, distance and particle primary. The energies were divided into 50 bins per decade, ranging from $10^{18}$ to $10^{22}$~eV. The distances were divided into 118 logarithmic bins, ranging from 3 to 3342 Mpc. Five representative primaries were considered: proton, helium, nitrogen, silicon, and iron. Since no arrival direction data is considered, a 1-D simulation in which no effect of magnetic fields is taken into account was performed. For each LIV coefficient value, $10^{4}$ events were simulated in each bin of the 3-D grid, resulting in a total of 118.59 million events for each LIV scenario. The information about the particle's energy and species at the source, $E_{s}$, $I_{s}$, the source distance $D_{s}$ and energy and species at Earth, $E_{E}$, $I_{E}$ was recorded.

The fiducial astrophysical model considers every source accelerating particles as a standard candle with an injected spectrum given by

\begin{equation}
    \label{eq:spectra}
    \left.\frac{dN}{dE_{s}}\right|_{i_s} = N_{0} f_{i_s} \left(\frac{E_s}{1 \, \rm{EeV}}\right)^{-\Gamma} e^{-E_s/(Z_i R_{\rm{max}})},
\end{equation}

where $i_s$ denotes the particle species at the source, $\Gamma$ the spectral index, $R_{\rm{\max}}$ the maximum rigidity at the sources, and $f_i$ and the initial primary fractions. The free fit parameters are $\Gamma$, $R_{\rm{max}}$ and $F_i$, the integrated individual contribution between 1 EeV and the maximum energy break for that primary ($f_i = F_i / (Z_i R_{\rm{max}})^{(\Gamma-1)}$). This is chosen as the fitting procedure is shown to be more stable with the change in definition since the relative differences in $F_i$ are usually smaller than those in $f_i$~\cite{deOliveira:2024juz}. The normalization, $N_0$, is taken such that the simulated flux matches the data for $E=10^{18.95}$~eV.

The $X_{\rm{max}}$ distributions are obtained by convoluting the Gumbel distributions~\citep{Gumbel} for each primary and energy arriving on Earth with the $X_{\rm{max}}$ acceptance given in Ref.~\citep{TACombinedFit}. The \texttt{EPOS-LHC} model~\citep{EPOS} was considered for the hadronic interactions. 

The combined fit was performed both by considering and not considering Amaterasu. For both the spectrum and $X_{\rm{max}}$ counts the likelihood for Poisson distributions was calculated and compared to the likelihood of a model that perfectly describes the data, resulting in the so-called deviance, which can be seen as a more generalized $\chi^2$ distribution. Seven free parameters\footnote{One of the primary fractions is fixed such that $\sum_i F_i = 1$.} and 16 (17) spectral and 55 $X_{\rm{max}}$ bins with non-zero counts were considered for the case without (with) Amaterasu, leading to 64 (65) number of degrees of freedom (NDF).

The systematics in the energy scale and in the $X_{\rm{max}}$ were investigated by performing the fit again with the measured data scaled by $\pm \sim 50\%$ and $\pm 100\%$ of the systematics of the Telescope Array experiment, i.e., $E \rightarrow E + (-22,15,0,15,22)\%$ and $X_{\rm{max}} \rightarrow X_{\rm{max}} \pm (-15,-7,0,7,15) \, \rm{g/cm^2}$, resulting in 2x25 total fits for each LIV parameter.

\subsection{Fit results}

For all the fits, the best deviance was found with the systematic shift of $E \rightarrow E - 22\%$ and $X_{\rm{max}} \rightarrow X_{\rm{max}} + 15 \, \rm{g/cm^2}$.

\begin{figure}[t]
\centering
\includegraphics[width=0.8\textwidth]{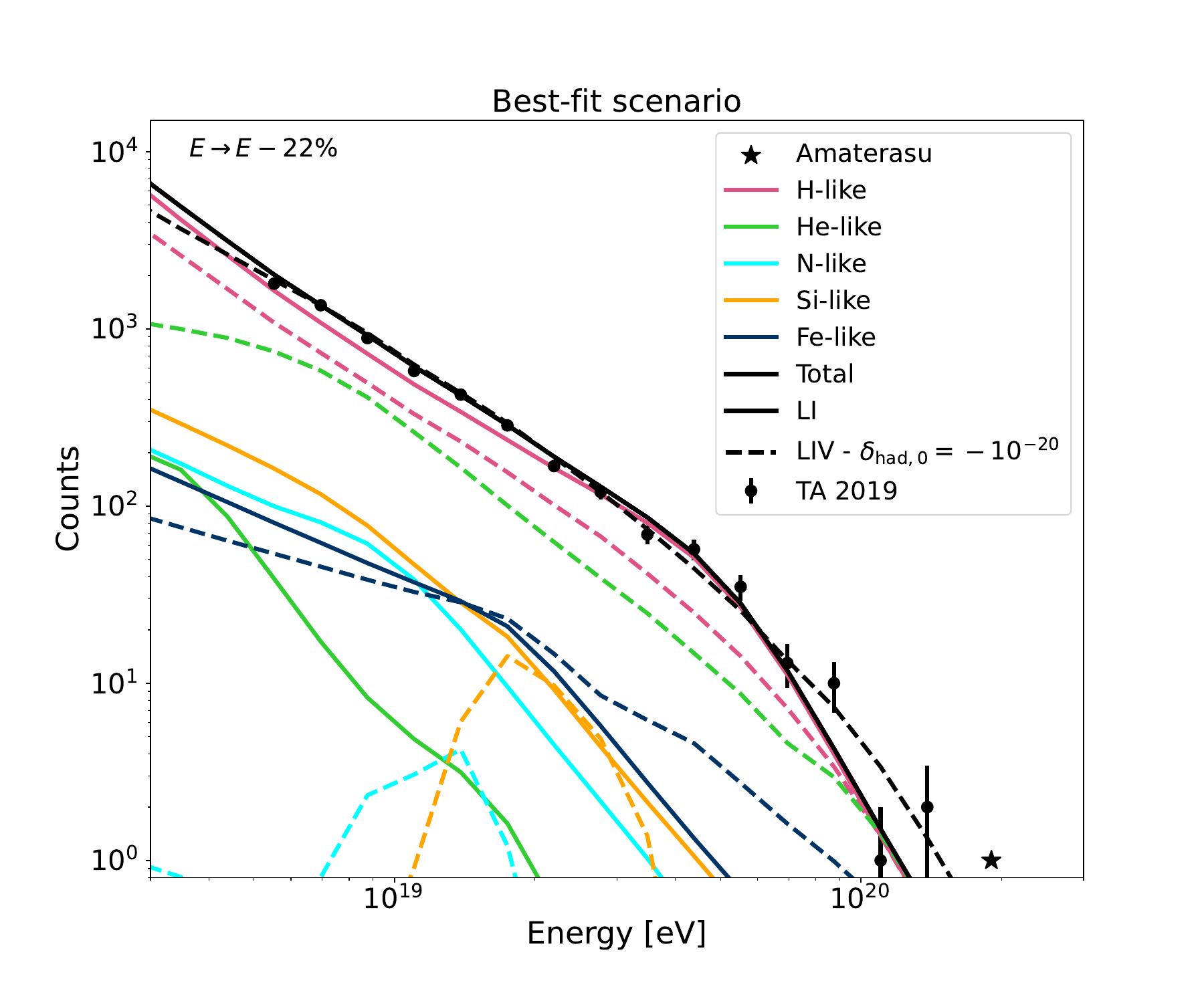}
\caption{\label{fig:Spectra}Spectra for the best fit. The different colors represent the contribution of different primaries at Earth and are divided into H-like ($A = 1$), He-like ($2 \leq A \leq 4$), N-like ($5 \leq A \leq 22$), Si-like ($23 \leq A \leq 38$), and Fe-like ($39 \leq A \leq 56$). The dashed and full lines are for the LI case and the LIV case with $\delta_{\rm{had,0}} = -10^{-20}$ respectively. For both cases, the best fit was found with a systematic shift of $E \rightarrow E-22\%$ and $X_{\rm{max}} \rightarrow X_{\rm{max}} + 15 \, \rm{g/cm^2}$.}
\end{figure}

\begin{figure}[t]
\centering
\includegraphics[width=0.8\textwidth]{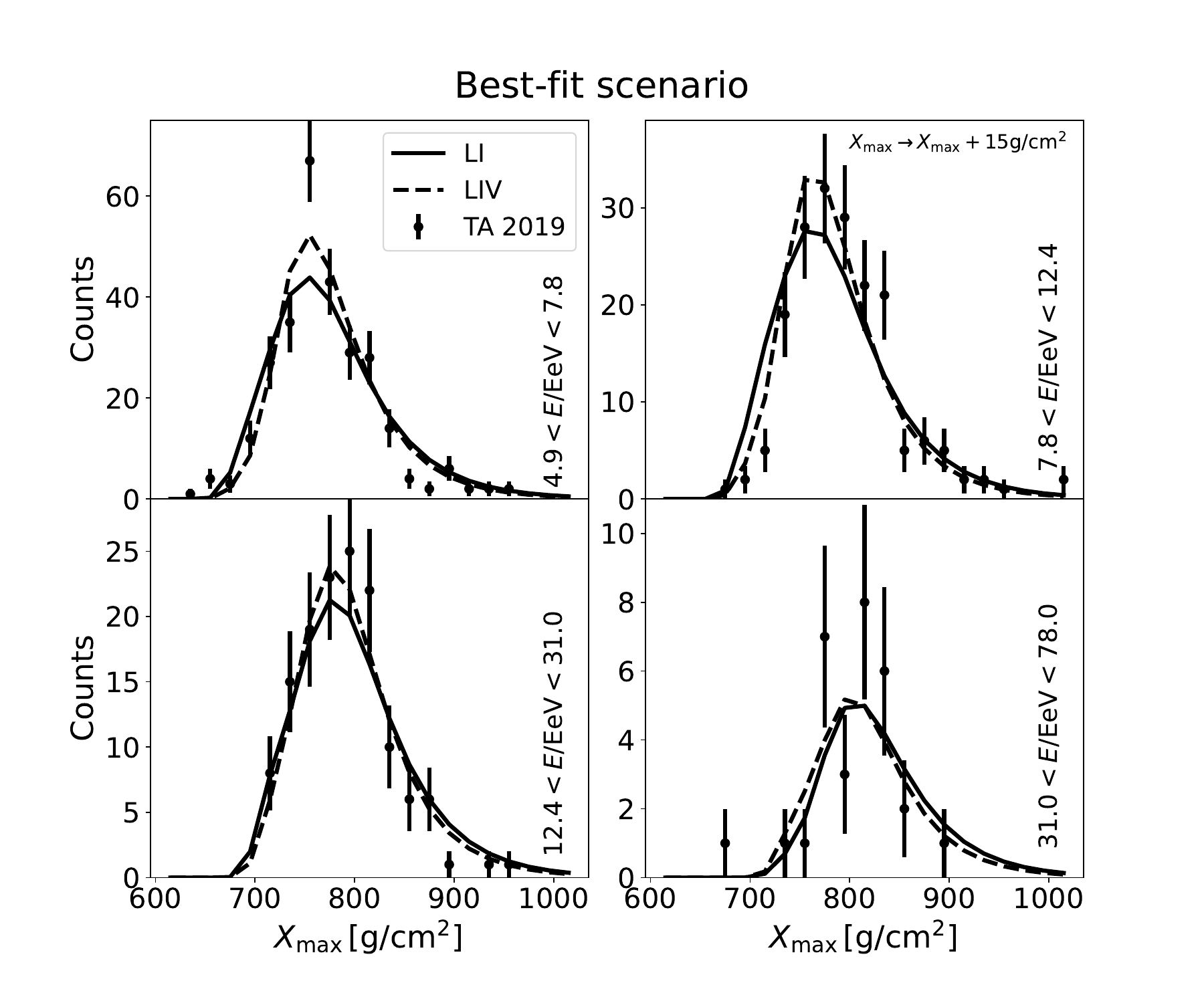}
\caption{\label{fig:Xmax}$X_{\rm{max}}$ distributions for the best fit. The dashed and full lines are for the LI case and the LIV case with $\delta_{\rm{had,0}} = -10^{-20}$ respectively. Each panel represents a different energy bin. For both cases, the best fit was found with a systematic shift of $E \rightarrow E-22\%$ and $X_{\rm{max}} \rightarrow X_{\rm{max}} + 15 \, \rm{g/cm^2}$.}
\end{figure}

Figure~\ref{fig:Spectra} and figure~\ref{fig:Xmax} show the best-fit spectra and $X_{\rm{max}}$ distributions on Earth in an LI scenario and in a LIV scenario with $\delta_{\rm{had},0} = -10^{-20}$, respectively. The recovery at the high-energy end of the spectrum is significant and comes from the fact that, as previously discussed, nuclei undergo fewer interactions. The intermediate and low energy ranges of the proton spectrum are very similar for both cases. This creates a scenario that still describes well the low-energy end of the spectrum, which dominates the statistics while improving the description of the highest energy events. The LI fit favors a proton-dominated scenario. When LIV is considered, a shift to a slightly heavier composition is seen due to a larger helium contribution. This happens due to less photodisintegration experienced by these particles.

The differences for the $X_{\rm{max}}$ are smaller since the fluorescence detector (FD) measurements do not reach the highest energies, at which LIV effects are more significant. The differences seen in the first two energy bins come from the larger contribution of helium in the LIV case.

\begin{figure}[t]
\centering
\includegraphics[width=0.8\textwidth]{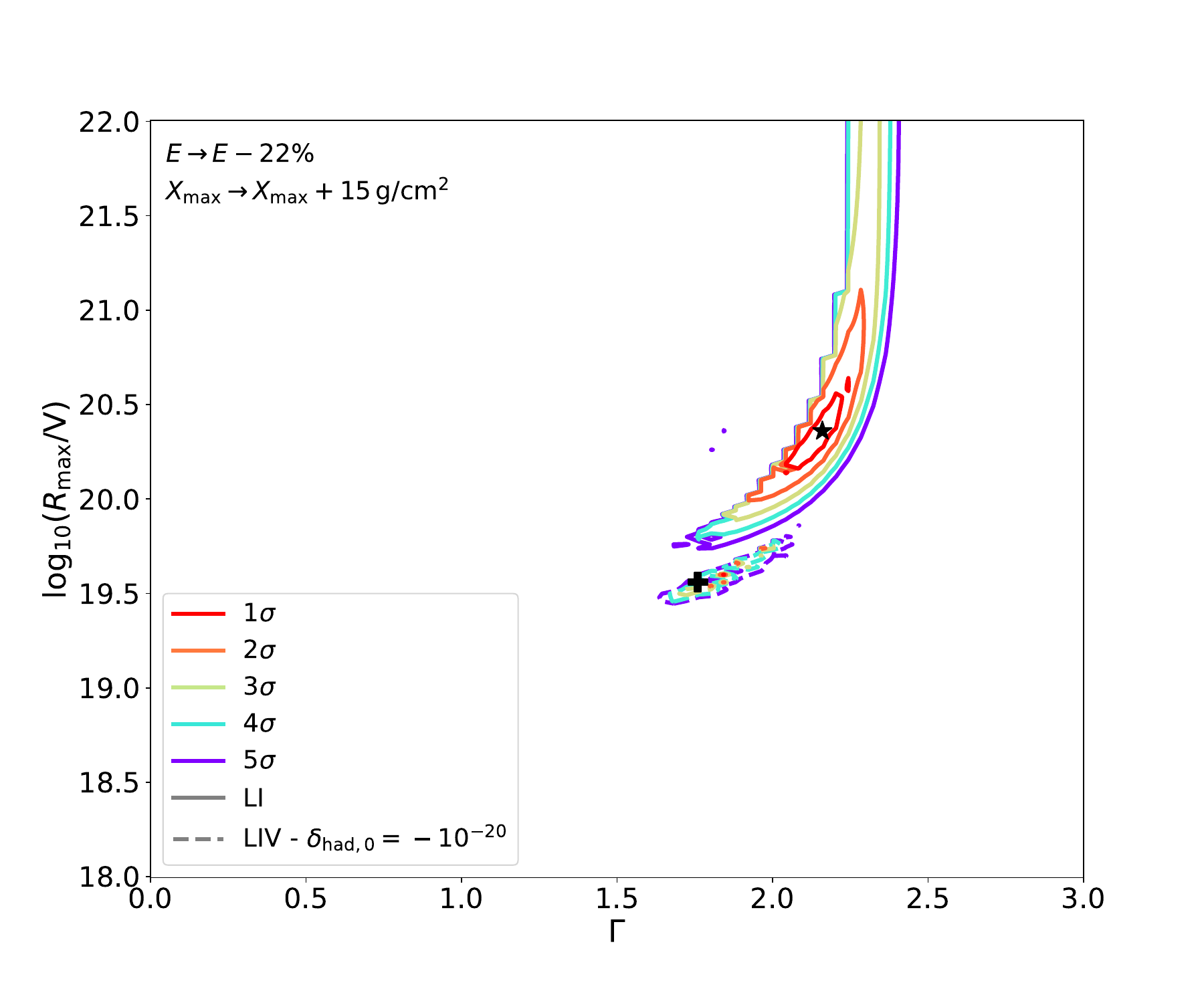}
\caption{\label{fig:Likelihood}Likelihood contour for the fit. The different colors show the contours for 1, 2, 3, 4, and $5\sigma$. The full and dashed lines are for the cases with LI and $\delta_{\rm{had,0}} = -10^{-20}$, respectively. The star and plus markers show the minimum for these respective cases. In every scenario, the best fit was found with a systematic shift of $E \rightarrow E-22\%$ and $X_{\rm{max}} \rightarrow X_{\rm{max}}+15 \, \rm{g/cm^2}$.}
\end{figure}

\begin{figure}[t]
\centering
\includegraphics[width=0.49\textwidth]{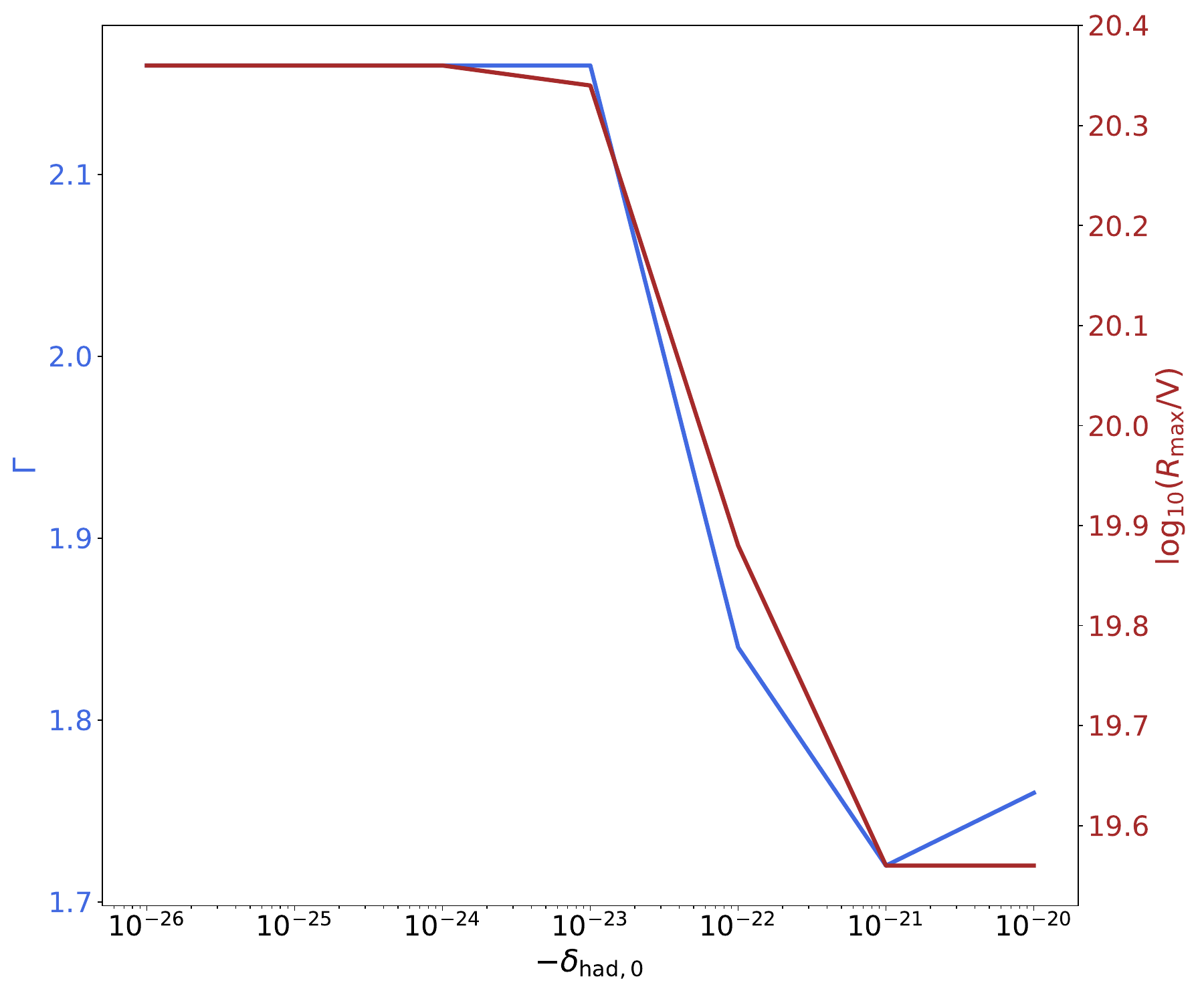}
\includegraphics[width=0.49\textwidth]{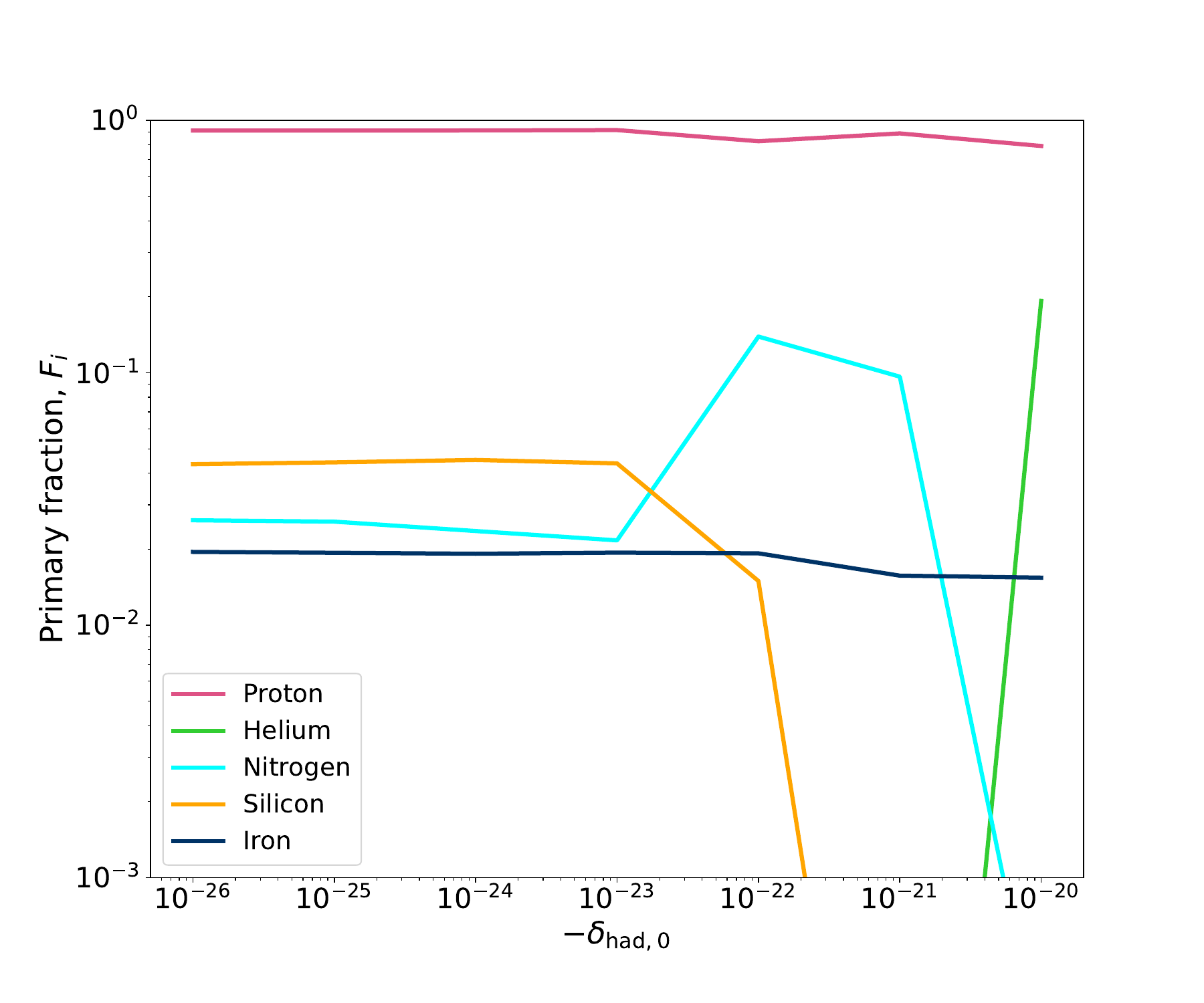}
\caption{\label{fig:FitParameters}Best-fit parameters as a function of the LIV coefficient. The left panel shows the spectral index (left axis) and maximum rigidity at the sources (right axis). The right panels show the integrated primary fractions, $F_i$, defined as the integrated primary fraction between 1 EeV and the maximum energy break for that primary.}
\end{figure}

Figure~\ref{fig:Likelihood} shows the likelihood contour for the fit of these two cases. The presence of LIV shifts the region of the phase space that best fits the data. This is best quantified in figure~\ref{fig:FitParameters}, in which the evolution of the fit parameters with the LIV coefficients is shown. In stronger LIV scenarios, the cutoff cannot be well described by the energy-dependent propagation horizon and, thus, a model with a lower maximum power of acceleration at the sources is favored. As already shown in previous combined fit works~\citep{AugerCombinedFit,AugerLIV}, the best spectral index is strongly correlated to the best rigidity cutoff. This behavior is also seen here, with the best-fit spectral index getting smaller for larger LIV coefficients. The primary fractions do not strongly vary, with a proton-dominated scenario being favored in all the considered scenarios. It is, though, noteworthy that the primary fraction as defined in this work refers to the integral fraction at the sources between 1 EeV and the maximum energy break. As the cutoff is rigidity (and not energy) dependent, for energies above $R_{\rm{max}}$, the injected composition becomes less proton-dominated and, thus, heavier. 

\begin{figure}[t]
\centering
\includegraphics[width=0.8\textwidth]{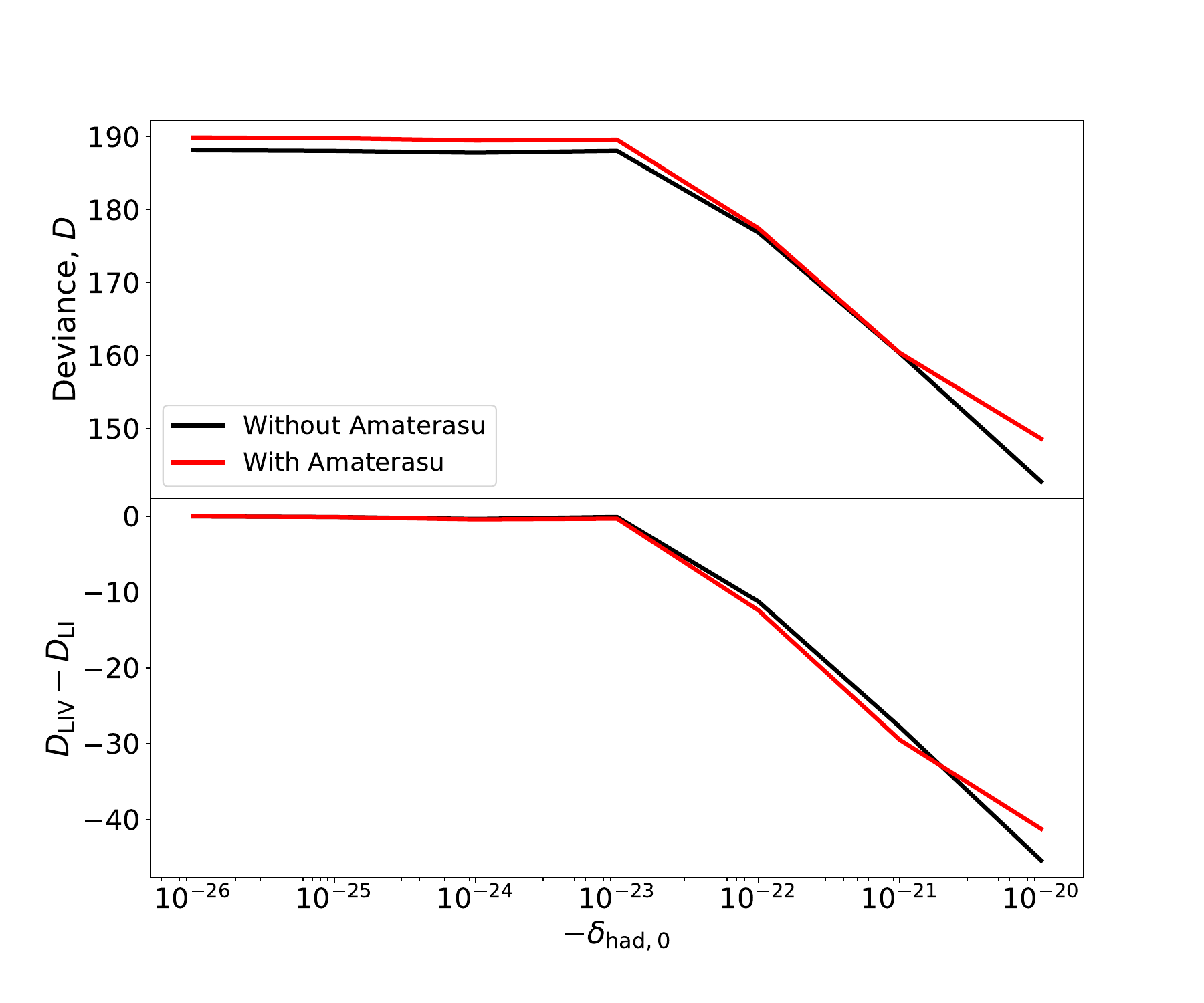}
\caption{\label{fig:Deviance}Deviance for the best fit found for each LIV coefficient. Black and red lines show the results not considering and considering the measurement of Amaterasu, respectively. The bottom panel shows the relative deviance to the LI case.}
\end{figure}

While figure~\ref{fig:Spectra} provides a visual indication of the improvement of the description of the data within LIV assumptions, a more quantitative measurement is needed. Figure~\ref{fig:Deviance} shows the evolution of the deviance with the LIV coefficient, both with and without considering Amaterasu. Even without Amaterasu, LIV scenarios are favored due to their better description of the high-energy end of the spectrum. With Amaterasu, the improvement of a LIV description becomes even more significant, highlighting the potential of new physics to describe this unusual event.

\section{Compatibility with results from the Pierre Auger Observatory}
\label{sec:auger}

The Amaterasu particle has been detected by the Telescope Array Experiment and, thus, a combined fit of the data of this experiment is the natural first step to check the robustness of the LIV scenarios here proposed. Nevertheless, the Pierre Auger Observatory has been taking data in the same energy range with an area about four times larger, leading to a larger exposure and smaller statistical uncertainties. A joint description of the data of both experiments is a great challenge due to different calibrations, regions of the sky being observed, energy scales, among others.

An excess at the highest energies is found for the Telescope Array spectrum even after exploring the ideal energy scale shifts~\cite{Tsunesada:2021qO}. Possible explanations include, e.g., a local source of UHECR~\cite{Plotko:2022urd}.

As for the $X_{\rm{max}}$, a more significant difference is found. The composition measured by Telescope Area seems to be proton dominated, while the composition measured by the Pierre Auger Observatory points towards a mixed and intermediate composition and rules out a proton-dominated scenario. A joint work between both collaborations indicates that the Telescope Array data are consistent also with the mixed composition measured by the Pierre Auger Observatory~\cite{deSouza:2017wgx}.

A combined fit of only the Pierre Auger data under LIV assumptions has already been investigated by the Pierre Auger Collaboration~\cite{AugerLIV}. A combined fit of the data of both experiments has proven to be quite challenging even if no new physics is assumed and is, thus, out of the scope of this work. Nevertheless, the results for the best-fit scenario obtained in the previous section point toward a proton-dominated composition and could, thus, be in contrast with data from the Pierre Auger Observatory. For that reason, we consider a second scenario with a mixed and intermediate composition, henceforth called ``intermediate composition scenario''. This is achieved by forcing the resulting $\left<X_{\rm{max}}\right>$ from the model to be lower (within systematics) than the $\left<X_{\rm{max}}\right>$ measured by the Pierre Auger Observatory~\cite{Yushkov:2020nhr} for energies above $10^{18.8}$~eV.

\begin{figure}[t]
\centering
\includegraphics[width=0.8\textwidth]{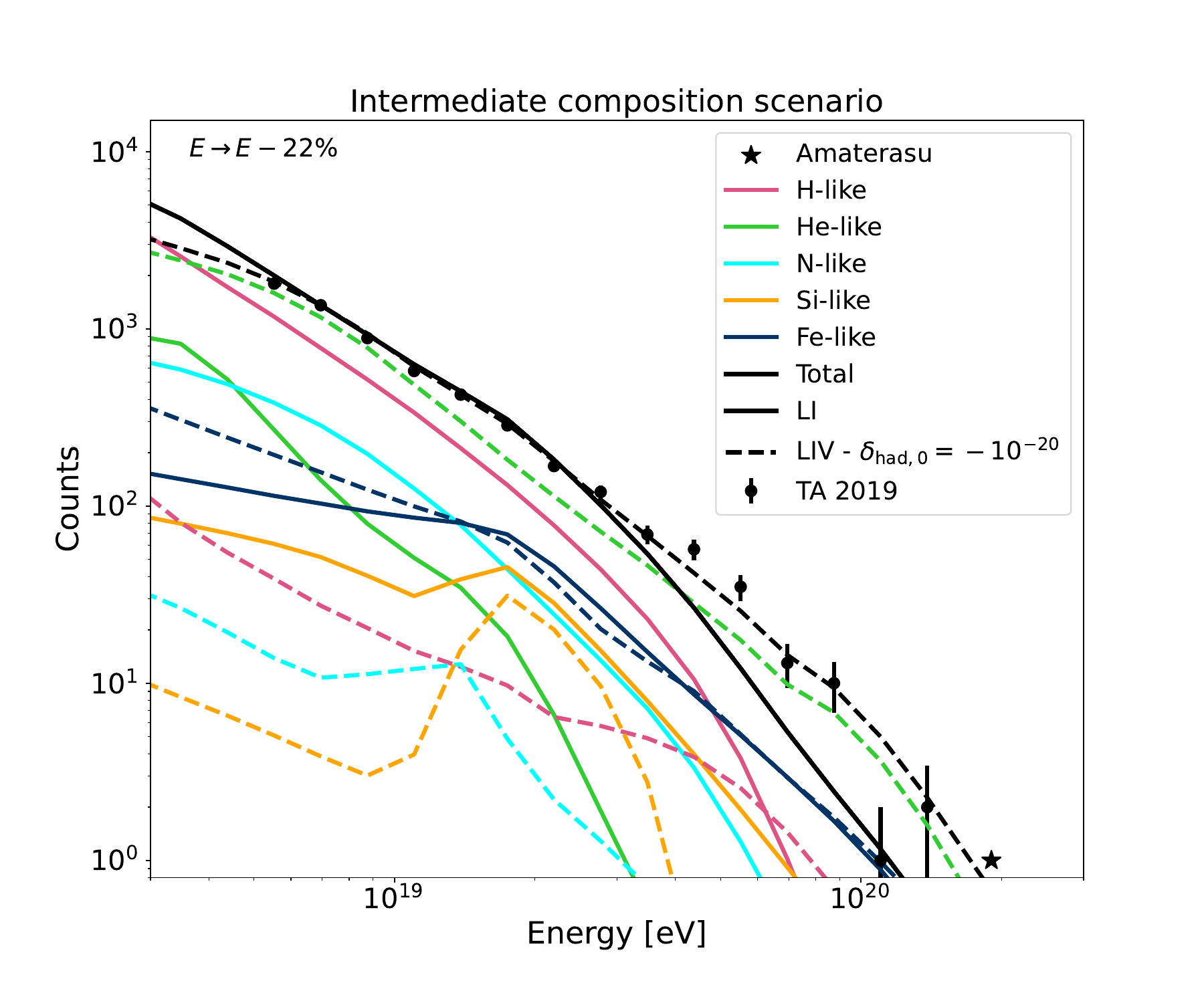}
\caption{\label{fig:Spectra-AugerScenario}Same as figure~\ref{fig:Spectra}, but for the intermediate composition scenario. The best fit was found with a systematic shift of $E \rightarrow E-22\%$ and no systematic shift on $X_{\rm{max}}$.}
\end{figure}

\begin{figure}[t]
\centering
\includegraphics[width=0.8\textwidth]{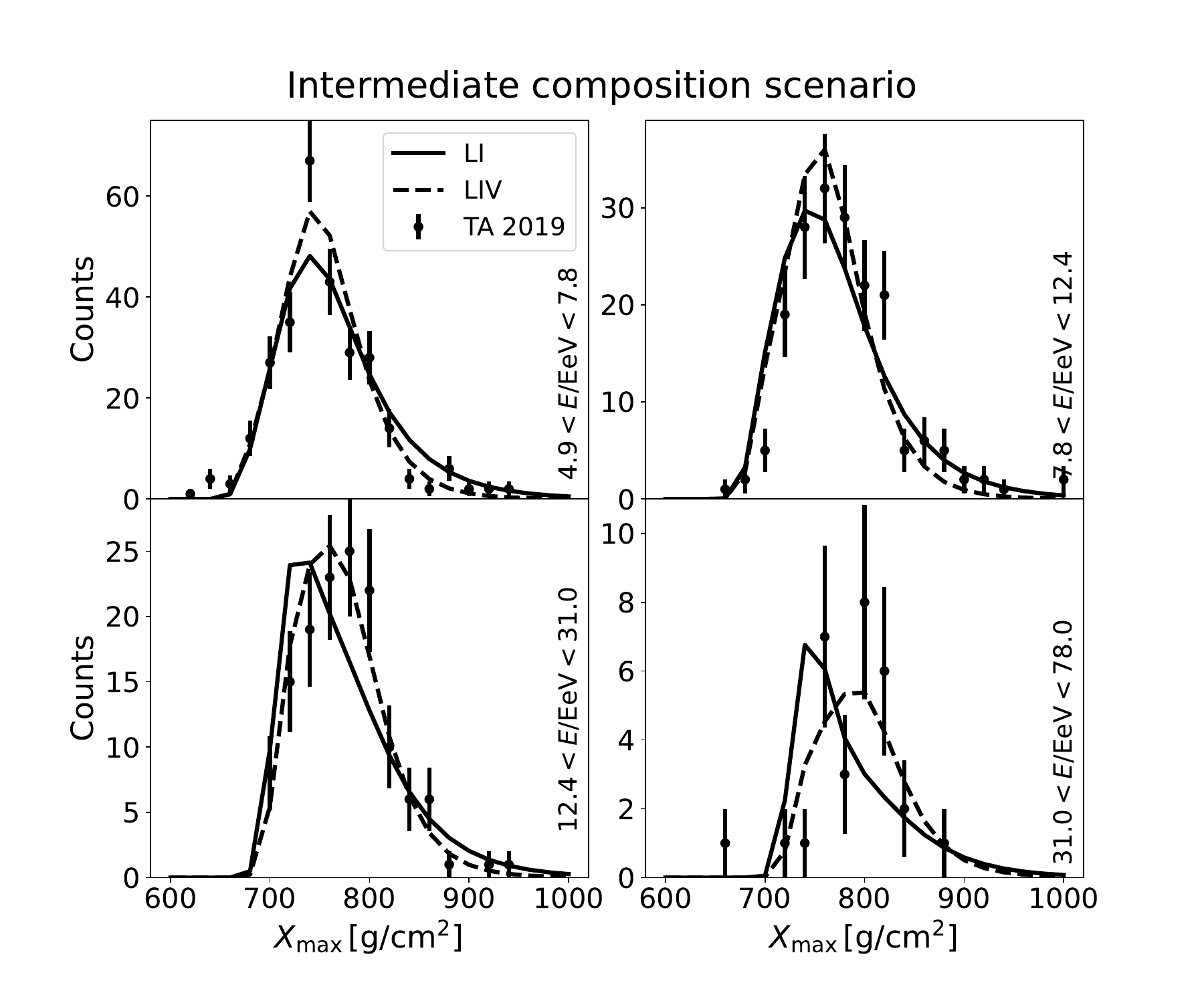}
\caption{\label{fig:Xmax-AugerScenario}Same as figure~\ref{fig:Xmax}, but for the intermediate composition scenario. The best fit was found with a systematic shift of $E \rightarrow E-22\%$ and no systematic shift on $X_{\rm{max}}$.}
\end{figure}

\begin{figure}[t]
\centering
\includegraphics[width=0.8\textwidth]{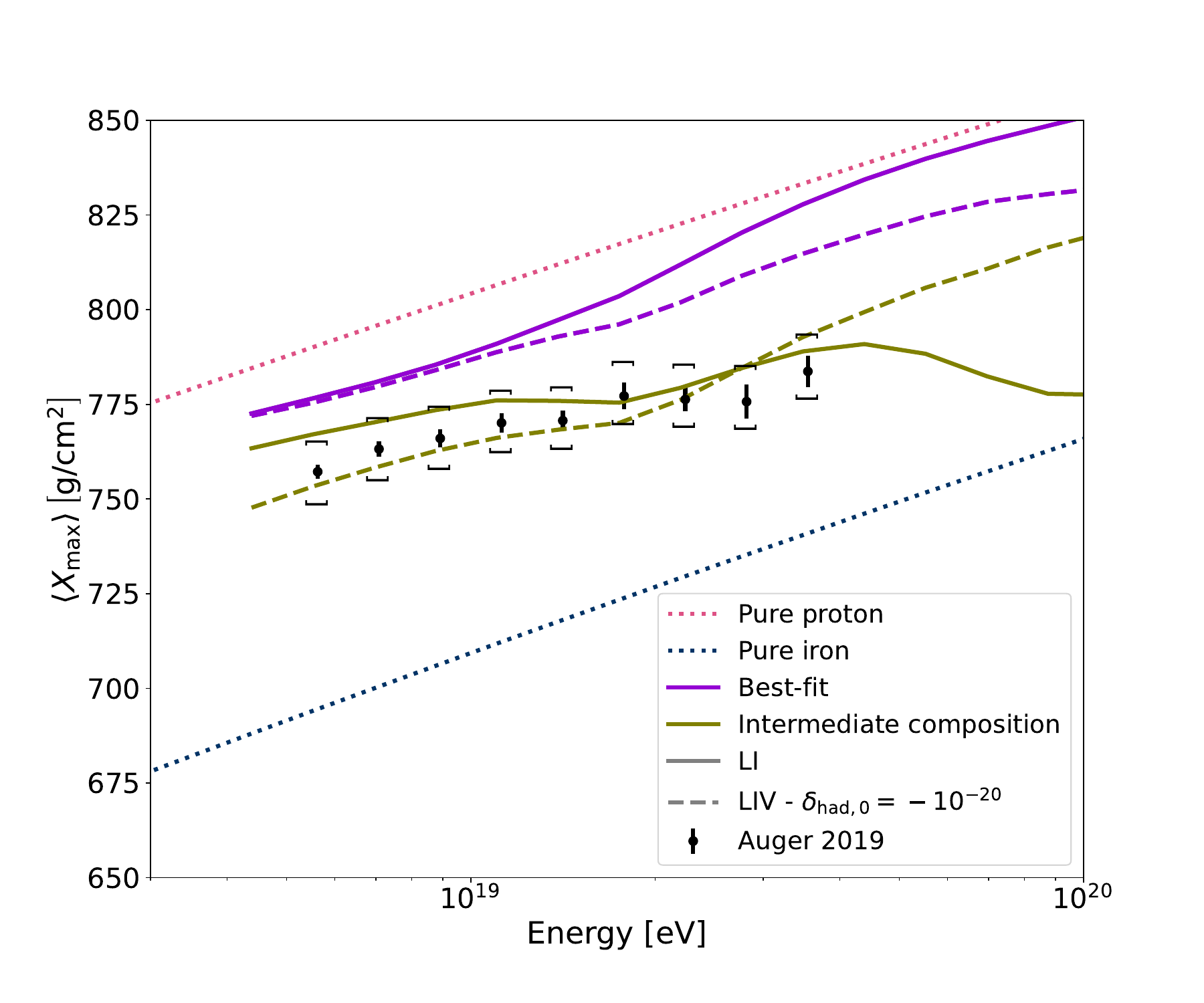}
\caption{\label{fig:comparexmax}Mean of the $X_{\rm{max}}$ distribution as a function of energy. The points show the data from the Pierre Auger Observatory~\cite{Yushkov:2020nhr} and the open brackets show the systematic uncertainties for these. The dotted pink and blue lines show the expected $\left<X_{\rm{max}}\right>$ for a pure proton and iron composition, respectively. The \texttt{EPOS-LHC} model was used for the hadronic interaction. The dark purple and olive lines show the predicted values for each scenario. For the intermediate composition scenario, the $\left<X_{\rm{max}}\right>$ was forced to be below the Pierre Auger data within systematics for $E > 10^{18.8}$~eV.}
\end{figure}

The same fitting procedure has been repeated for this scenario. The best fit was achieved with no systematic shift in the $X_{\rm{max}}$ distribution and an energy scale of $E \rightarrow E - 22\%$. Figures~\ref{fig:Spectra-AugerScenario} and \ref{fig:Xmax-AugerScenario} show the resulting spectra and $X_{\rm{max}}$ for LI and LIV with $\delta_{\rm{had},0} = - 10^{-20}$, while figure~\ref{fig:comparexmax} shows the comparison of the predicted $\left<X_{\rm{max}}\right>$ to the data from the Pierre Auger Observatory. As expected, a much larger contribution of heavier nuclei is seen. For the LI case, the suppression cannot be well described, leading to a bad agreement to data in the highest energies. This is significantly improved in the LIV scenario, with many more UHECR arriving at the highest energies due to the lack of interactions during propagation. The composition is remarkably changed even in the lowest energies, getting heavier and almost fully helium dominated for the LIV scenario.

\begin{figure}[t]
\centering
\includegraphics[width=0.8\textwidth]{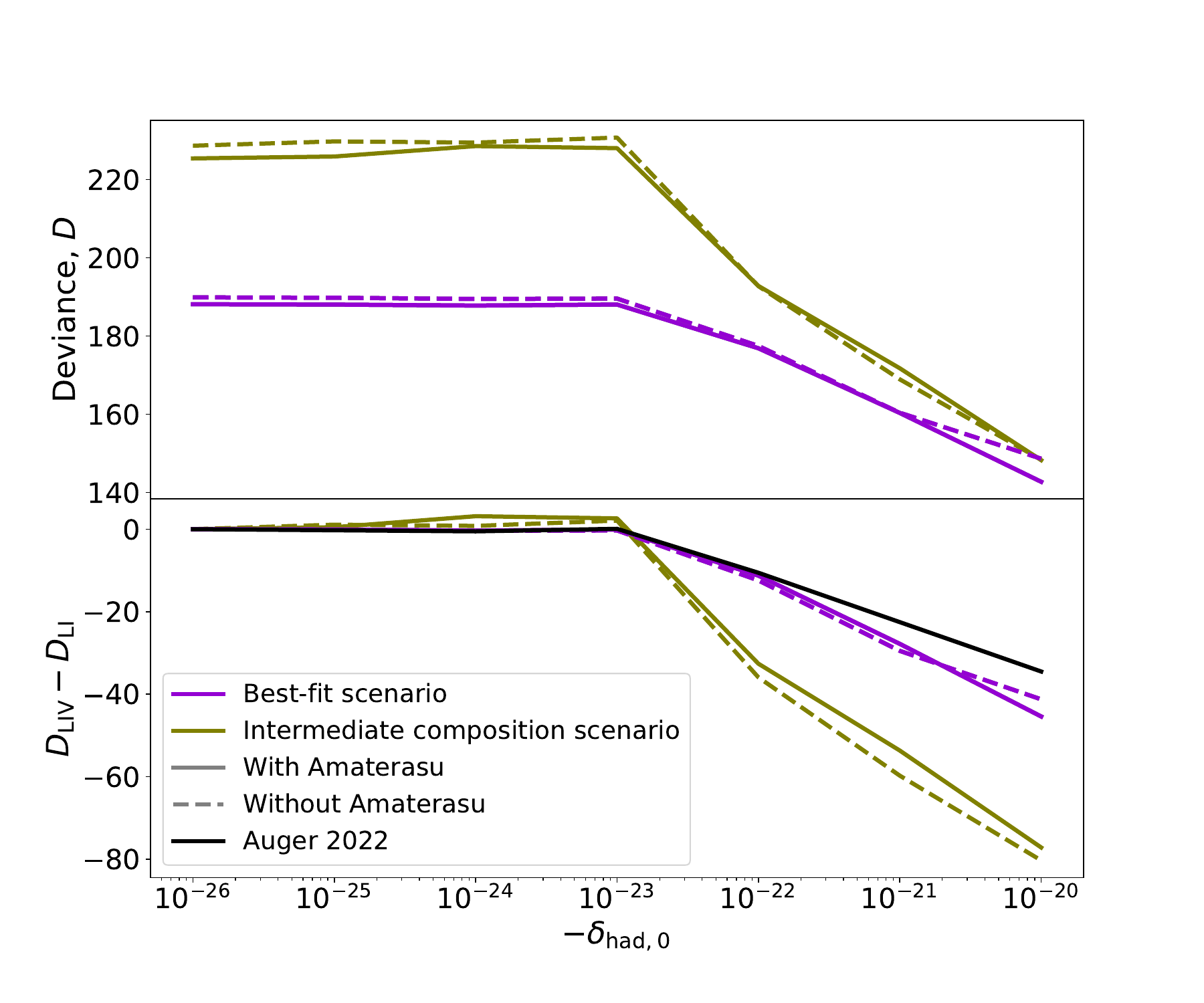}
\caption{\label{fig:Deviance-AugerScenario}Deviance versus LIV coefficient for best-fit scenario (dark purple) and intermediate composition scenario (olive). The full and dashed lines represent the fits with and without Amaterasu, respectively. The bottom panel shows the relative difference to the LI case. The black line is taken from Ref.~\cite{AugerLIV} and shows the relative deviance found by the Pierre Auger Observatory in their similar study.}
\end{figure}

The clearly seen improvement in the LIV case is quantified in figure~\ref{fig:Deviance-AugerScenario}, which shows the deviance as a function of the LIV coefficient for every scenario considered. As expected, the total deviance for the LI case is much worse in the intermediate composition scenario in relation to the best-fit scenario. This reinforces the case that a proton-dominated scenario better describes the Telescope Array data in an LI scenario. Nevertheless, when LIV is considered the improvement in the description of the data is larger for the intermediate composition scenario and similar values for the deviance as those from the best-fit scenario are reached for $\delta_{\rm{had},0} < -10^{-21}$. This shows that the improvement seen in a LIV description is also consistent with an intermediate composition compatible with the one from the Pierre Auger Observatory. But also that LIV allows the data from the Telescope Array Experiment to be described with an intermediate composition as well as it is described with a proton-dominated one, loosening some tension on the differences between the $X_{\rm{max}}$ distributions measured by the two experiments.

The relative deviance of a similar LIV study from the Pierre Auger Collaboration~\cite{AugerLIV} is also shown and the behavior agrees very well with the relative deviances from this work. No significant differences are seen up to $\delta_{\rm{had},0} = - 10^{-23}$, above which an improvement in the description in the data is seen. This confirms that the LIV scenarios preferred by this analysis are consistent not only with Telescope Array data, but also with Pierre Auger data.

\section{Conclusions\label{sec:conclusions}}

In this work, we explored the potential of describing the intriguing measurement of the Amaterasu particle under a LIV scenario.

At first, the kinematics of propagation interactions of UHECR under a phenomenological approach to LIV in the hadronic sector were calculated. The main effect for $\delta_{\rm{had},0} < 0$ is a significant increase of the mean free path above an energy that is dictated by the LIV coefficient. The increase in the total mean free path for the energy of Amaterasu becomes significant for $-\delta_{\rm{had},0} > 10^{-22}$. Even for the least conservative case, iron, the total mean free path can increase from a few Mpc to hundreds of Mpc. This would soften the tension with the origin of Amaterasu since no nearby (up to few tens of Mpc) strong source in the local void would be required to have accelerated it. This is robust within systematic uncertainties of the energy of Amaterasu.

Nevertheless, a LIV scenario that softens that tension still must be consistent with the rest of the data. For that reason, we performed a combined fit of the spectrum and composition data of Telescope Array for each LIV coefficient considered. The systematic uncertainties in the energy and $X_{\rm{max}}$ were taken into account by shifting the measured data by $E \rightarrow E \pm 22\%$ and $X_{\rm{max}} \rightarrow X_{\rm{max}} \pm 15 \, \rm{g/cm^2}$. For every fit performed, the deviance was the lowest for $E \rightarrow E - 22\%$ and $X_{\rm{max}} \rightarrow X_{\rm{max}} + 15 \, \rm{g/cm^2}$.

The LIV best-fit spectra describe the low and intermediate energy range of the measured energy spectrum as well as the LI case while improving the description of the highest energy events, increasing the probability of detecting an event such as Amaterasu. A lower rigidity cutoff is preferred by strong LIV scenarios, meaning that the spectrum cutoff is dominated by the power of acceleration of the sources, since the propagation horizon is significantly increased.

Even without Amaterasu, the deviance of the best fit is significantly improved under LIV assumptions. With Amatarasu, this improvement is even larger, confirming that a LIV model in which the tension with the origin of Amaterasu is softened is still consistent with the rest of the data from the Telescope Array.

As a LIV scenario could only be robust if also compatible with data from other experiments, we further explored its compatibility with results from the Pierre Auger Collaboration. A combined fit of the data of both experiments together is extremely complex and a subject of discussion even for scenarios without new physics and, thus, out of the scope of this work. The main incompatibility with the Pierre Auger data seen in the best-fit scenarios of this work is the proton-dominated composition. For that reason, we explored the results when an intermediate composition is assumed. This was done by forcing the $\left<X_{\rm{max}}\right>$ to be no larger (within systematics) than the one measured by the Pierre Auger Observatory. Similar findings are seen, with LIV leading to an improved description of the data, showing that the LIV improvement is also consistent with an intermediate composition compatible with that from the Pierre Auger Observatory.

The deviance for the intermediate composition scenario reaches deviance values similar to those of the proton-dominated scenario when LIV is considered. This shows that, if LIV is considered, the data from Telescope Array can be described with an intermediate composition as well as it is described with a proton-dominated one. This could loosen the tension on the difference between the $X_{\rm{max}}$ distributions of the experiments.

The results of the Pierre Auger Collaboration in their investigation of LIV effects in their data~\citep{AugerLIV} are also compatible with the results of this work. In both scenarios here considered and in the results from the Pierre Auger Collaboration, the deviance does not change significantly up $\delta_{\rm{had},0} = -10^{-22}$, above which the data is best described.

With that, LIV is raised as a robust possible explanation for the Amaterasu particle, while still being consistent with (and even improving) the description of the data of both the Telescope Array Experiment and the Pierre Auger Observatory and relieving the tension on the differences on the $\left<X_{\rm{max}}\right>$ measured by both experiments.

It must, however, be noted that explanations within standard physics cannot be ruled out. The Amaterasu particle could be described by, e.g., transient events or ultra-heavy UHECR, while the combined fit could be significantly improved with a less simplistic astrophysical model with, e.g., a larger relative contribution of local sources located within a few tens of Mpc.

\acknowledgments
RGL thanks the referee for valuable feedback that lead to the improvement of the manuscript.

\bibliographystyle{JHEP}
\bibliography{biblio.bib}
\end{document}